\documentclass{sigchi}

\usepackage[usenames]{color}

\usepackage{times}
\usepackage{url}
\usepackage{graphics}
\usepackage[pdftex]{hyperref}
\usepackage{subfigure}
\usepackage[english]{babel}
\usepackage{recycle}
\usepackage{subfigure}
\usepackage{cite}
\usepackage{url}
\usepackage{color}
\usepackage{pifont}
\usepackage{amsmath,amssymb,amsfonts}
\hypersetup{%
pdftitle={Friends FTW! \\ Friendship, collaboration and competition in Halo:\! Reach},
pdfauthor={Winter Mason and Aaron Clauset},
pdfkeywords={Social Mining, Group Dynamics, MMO},
bookmarksnumbered,
pdfstartview={FitH},
colorlinks,
citecolor=black,
filecolor=black,
linkcolor=black,
urlcolor=black,
breaklinks=true,
}
\newcommand{\comment}[1]{}
\definecolor{Orange}{rgb}{1,0.5,0}

\begin{document}
% to make various LaTeX processors do the right thing with page size

\setlength{\paperheight}{11in}
\setlength{\paperwidth}{8.5in}
\setlength{\pdfpageheight}{\paperheight}
\setlength{\pdfpagewidth}{\paperwidth}

\title{Friends FTW! \\ Friendship, Collaboration and Competition in Halo:\! Reach}

\numberofauthors{2} 
\author{
% 1st. author
\alignauthor Winter Mason \\
       \affaddr{Stevens Institute of Technology}\\
       \email{m@winteram.com}
% 2nd. author
\alignauthor Aaron Clauset \\
       \affaddr{University of Colorado, Boulder}\\
       \affaddr{Santa Fe Institute}\\
       \email{aaron.clauset@colorado.edu}
}

\maketitle

\begin{abstract}

How important are friendships in determining success by individuals and
teams in complex collaborative environments? By combining a novel data set
containing the dynamics of millions of \emph{ad hoc} teams from the popular
multiplayer online first person shooter \emph{Halo:\!\! Reach} with survey
data on player demographics, play style, psychometrics and friendships
derived from an anonymous online survey, we investigate the impact of
friendship on collaborative and competitive performance. In addition to
finding significant differences in player behavior across these variables,
we find that friendships exert a strong influence, leading to both
improved individual and team performance---even after controlling for the
overall expertise of the team---and increased pro-social behaviors. Players
also structure their in-game activities around social opportunities, and as
a result hidden friendship ties can be accurately inferred directly from
behavioral time series. Virtual environments that enable such friendship
effects will thus likely see improved collaboration and
competition. 

\end{abstract}




\section{Introduction}\label{sec:intro}

Given the ubiquity of friendships in social interactions and complex social
systems, the value of any particular friendship can be difficult to
quantify. But some things are known: friendships are useful for finding new
jobs, as in the ``weak ties'' or Granovetter effect~\cite{Granovetter73};
they are useful for marketing because personal attributes---including fixed
variables like age, ethnicity and location, as well as more fluid variables
like preferences and opinions---exhibit
homophily~\cite{shrum:cheek:hunter:1988,mcpherson:smith-lovin:cook:2001}
and thus correlate across friendship ties; and, they are useful for social
filtering or search~\cite{borgatti:cross:2003,bao:etal:2007}, i.e., your
friends are good at predicting what information you will like.

But, how valuable are friendships within complex collaborative
environments, particularly those found online? On the one hand, teams
composed of friends may perform better as a result of their extensive
collaboration history, i.e., friendships may increase awareness and
understanding of others' goals and
motivations~\cite{rittenbruch:mcewan:2009} or they may increase commitment
to group objectives~\cite{jehn:shah:1997}, thereby facilitating more
successful collaboration. Friendships may also reduce within-group
conflict~\cite{jehn:mannix:2001}, yielding similar results. On the other
hand, friendships could detract from performance because friends spend less
time focusing on group objectives and more time
socializing~\cite{shah:jehn:1993}.

Friendship may even be irrelevant for effective
collaboration.  If a competition's outcome depends mainly on team
coordination, teams composed of experts may coordinate effectively
regardless of within-team friendships: highly practiced individuals may
simply know from experience how to work well with other experts and
thus naturally anticipate or adapt to their teammates' actions (e.g., by
fitting into established and effective team roles). In such a
case, friendships do not matter and teams composed of
skilled strangers will perform best~\cite{shah:jehn:1993}. Thus, the extent
to which expertise and friendship matter to success is an open question in
complex collaborative environments, particularly those found online.

Here, we focus on the topic of friendships, collaboration and competition in online game
environments, and specifically in the first-person shooter (FPS) genre, in which
teams of players compete against each other in non-persistent
virtual worlds. We are also interested in the variety of social dynamics
naturally observed in this complex environment, and the extent to which
they vary with demographic and psychometric variables.  There are many
studies in CSCW on the design of, and human action within, multi-user
dungeons~\cite{muramatsu:ackerman:1998}, massively multiplayer online games
(MMOGs)~\cite{bardzell:bardzell:2008} (among others), social virtual
worlds~\cite{brown:bell:2006}, and non-persistent game
sites~\cite{mcewan:2012}.  Work on friendships in online gaming
has focused on how friendships and social relationships are a motivation to
play games~\cite{chen:2008, steinkuehler:williams:2006} or how playing
games builds offline relationships and social capital of the
players~\cite{jakobsson:taylor:2003, taylor:2003}.

Work on guilds and parties~\cite{williams2006n} in massively multiplayer
online role-playing games (MMORPGs) suggests that teams are important for
individual satisfaction and performance, and anecdotal evidence suggests
they are conducive to the formation of friendships. However, the impact of
friendships themselves on team or individual performance has not been
studied quantitatively (but see~\cite{jakobsson:taylor:2003} for some
qualitative insights). And, work on first-person shooter type games has
generally focused on qualitative analysis of user actions and in-game
communication~\cite{wright:boria:breidenbach:2002,manninen:kujanpaa:2005,reeves:brown:laurier:2009}.

Most work on the utility of friendships in teams is found in the
management literature. Several such studies argue that friendships play a significant
role in mitigating certain types of within-group conflicts; thus, teams of friends perform better because internal conflict handicaps performance~\cite{jehn:mannix:2001,
  shah:jehn:1993,jehn:shah:1997}.  Others have focused on
designing effective teams in business or educational
environments~\cite{ruef:aldrich:carter:2003, baldwin:etal:1997,
  balkundi:harrison:2006}. In these settings, friendships seem to improve
group cohesion, satisfaction, and some measures of performance. Thus, we  
expect friendships to play a significant role in the
collaboration effectiveness in virtual environments, including those found in
the team-based competitive environments of the FPS genre.

Finally, the increasing CSCW interest in ``serious games''~\cite{michael:chen:2005} as
a model for creating effective work spaces makes the study of real online
games, and particularly those that allow collaboration with both friends
and strangers, a valuable laboratory by which to understand the dynamics
and determinants of collaboration and sociality in virtual environments.

Here, we study these questions in the context of the popular online FPS game \emph{Halo:\!\! Reach}. We combine two large and novel data
sets: one composed of rich online behavioral samples from the events within
\textit{Reach} competitions and one composed of rich demographic, play style,
psychometric and friendship variables collected from players of
\emph{Reach} through an anonymous online survey. This combination of
in-game behavioral data with survey data on the same participants has been
used productively to study human behavior in
MMORPGs~\cite{williams:poole:contractor:2010,
  yee:ducheneaut:nelson:likarish:2011} and online parlor
games~\cite{mcewan:2012}, but this study is the first time it has been
applied to the FPS genre or to the question of friendship, collaboration
and competition.

The survey asked participants to label their online and offline friendships
with other \textit{Reach} players, and these data allow us to test the
hypothesis that friendships have a direct and significant impact on the
results of these competitions. Features within the game allow us to control
for the impact of expertise. An in-game ``matchmaking'' system draws groups
of friends and individual players from the general population to create
competitions; thus, players collaborate with and compete against a wide
variety of teammates and competitors. Finally, we note that the sheer size
of \textit{Reach}, which is played by millions of individuals, provides a
wealth of data by which to investigate these and other questions about
collaboration and competition.

We find that friendships significantly improve both individual and team
performance in these complex collaborative environments. Friendships are
not as important to success as the raw expertise of teammates (friendship
does not compensate for being a bad player),
but individuals themselves perform better the more friends they play with,
independent of their team's performance. Our survey shows that {\em Reach}
players are also highly social, tending to prefer team-oriented play and
experiencing a strong sense of group cohesiveness and
coordination. Furthermore, older players (24 or older; 30\% of players)
tend to be more socially oriented and exhibit greater proclivity for
pro-social behaviors. Above and beyond performance, friendships also
reshape the style of play within a competition, and players structure their
choices around opportunities to play repeatedly with friends. As a result,
friendships can be accurately inferred directly from behavioral time seres
data. These results illustrate the strong role that friendships play in
complex collaborative and competitive virtual environments. They also shed
light on social dynamics within FPS games and suggest algorithms designed
to account for the benefits of friendship in predicting or designing
competitive environments.

\section{Methodology} \label{sec:methods}

\subsection{Game Mechanics and Data}
\label{sec:mechanics:and:data}

{\em Halo:~Reach} is a multiplayer online first person shooter 
game played by more than ten million people worldwide.
It was publicly released by Bungie Inc., a former subdivision of Microsoft
Game Studios, on 14 September 2010 and {\em Reach} players have now
generated more than 1 billion games. Players choose from among seven game
types and numerous subtypes, which are played on more than 33 terrain maps
with 74 weapons. Games can be played alone, with or against other players
via the Xbox Live online system. Both individual game and player summaries
are available through the Halo Reach Stats API. Through this interface, we
collected the player details of all individuals who participated in our
online survey and each of their full game histories, which yielded
2,445,617 complete games.

Among other information, each game file includes the sequence of the
scoring events at the per-second resolution and a list of players by
team. Scoring events are annotated with the name of the player generating
the event (a unique Xbox Live gamertag), the number of points scored and
the player giving up the points (if applicable).

Unlike professional sports~\cite{gabel:redner:2011}, teams in {\em Reach}
do not persist across competitions. Instead, each time a competition is
created, individuals or small ``parties'' of players (typically friends)
are grouped into teams by an in-game ``matchmaking'' algorithm; when the
competition is complete, the players (or a subset) may choose to play
together again as a party, or may reenter the matchmaking process to find
new teammates or competitors. The result is that players routinely
collaborate with or compete against different strangers in successive
games. {\em Reach}'s matchmaking system uses an algorithm called
TrueSkill~\cite{herbrich:minka:graepel:2007} to ensure equal total
``skill'' for both teams, which allows us to control for this variable in our
subsequent analysis.

Among the sampled games, there are three basic types: \emph{campaign
  games}, a sequence of story-driven, non-competitive,
player-versus-environment (PvE) maps that many players complete prior to
trying other types of games; \emph{firefight games} (also PvE), in which a
team of human-controlled players battle successive waves of
computer-controlled enemies; and \emph{competitive games}, a
player-versus-player (PvP) game type, in which teams of equal size
(typically 2, 4, 6 or 8 players) compete to either be the first to reach
some fixed number of points or have the largest score after a fixed length
of time. (The precise number of players per team, points required to win
and length of a game depends on the game subtype.) In competitive games,
players' avatars move through the game map simultaneously, in real-time,
navigating complex terrain, acquiring avatar modifications and encountering
opponents. Points are scored by dealing sufficient damage to eliminate an
opposing avatar and for each such success, a team gains a single
point. Eliminated players must then wait several seconds before their
avatar is placed back into the game at one of several specified ``spawn''
locations.

\begin{figure}[t!]
\begin{center}
\includegraphics[width=0.95\columnwidth]{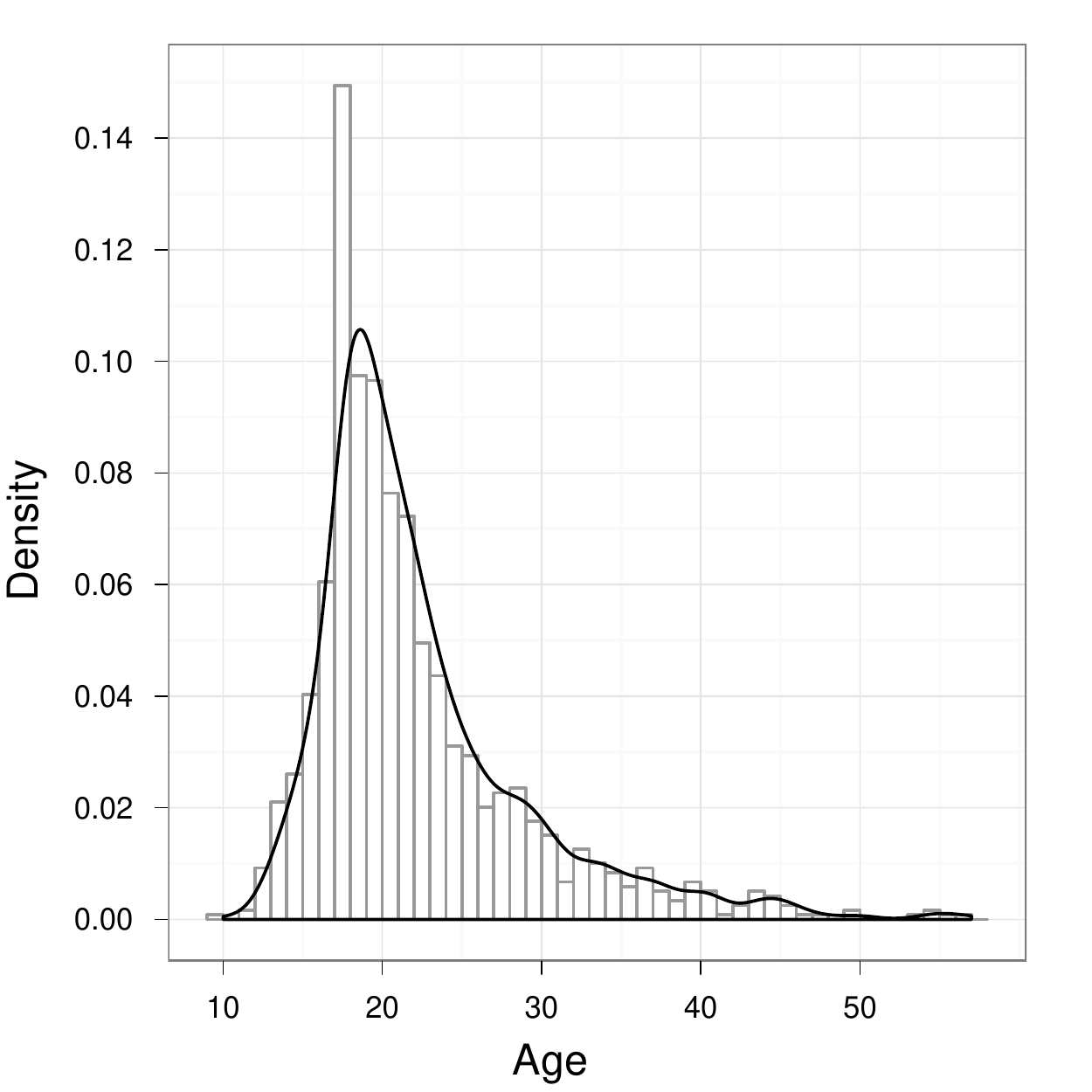}
\end{center}
\vspace*{-3mm}
\caption{Age distribution of the survey respondents.
\label{fig:age_dist}}
\vspace*{-3mm}
\end{figure}

\subsection{Survey Data}
In addition to the in-game behavioral data available through the API, we
collected data on 1191 unique {\em Reach} players via an anonymous
online survey.%
\footnote{The survey was open until November 2012, when Bungie, Inc.~turned
  off the API.  The anonymous design did not permit detailed interviews.} 
The survey design incorporated extensive feedback from expert
\textit{Reach} players, including ourselves, to better frame questions in
appropriate language and to focus on relevant social dynamics.

We advertised the survey through Halo-centric online forums and other
community websites. While this did result in a biased sample (see
below), the bias was towards participants who were active
and therefore had more in-game data to draw upon.  Participants supplied
their Xbox Live gamertag, which we used to download their {\em Reach}
player details and their entire {\em Reach} game history up to the date of
survey participation. Within the survey, an initial screening was conducted
based on participant age: participants who indicated their age was under 18
were routed through an additional email-based parental consent step.

The player detail files contain only summary statistics about {\em Reach}
game play and do not contain demographic or social network information. The
survey solicited these missing demographic variables (age, sex, location,
highest education level, primary language), along with preferred play style (lone
wolf, team leader, or team support), and psychometric responses like
measures of group cohesion~\cite{Stokes.1983},
entativity~\cite{McConnell.1997}, and conflict~\cite{Rahim.1983}. In
answering these questions, respondents were instructed to consider their
``primary Halo group,'' defined as the group of friends with whom they
primarily play {\em Reach} at the time of the survey. Finally, respondents
indicated their relationship (online friend, offline friend or not a
friend) with every other unique gametag that appeared in their game
history. Because this list could include hundreds or even thousands of
gamertags, it was sorted by the frequency of co-plays, which placed the most
likely true positives at the top.  The interpretation of the meaning of an
online and offline friend was left to the participant, although we
suggested online friends were people ``you play with regularly and would
say you know at least casually.''  This is conceptually distinct from any
feature available on Halo or XBox for indicating friendships.

\section{Player Attributes} \label{sec:survey}

\subsection{Demographics} \label{sec:demogs}

As is typical with many online games~\cite{williams2008plays} and
especially first person shooters~\cite{jansz:tanis:2007}, our survey
respondents were mostly young men.  The median age of our participants was
20, which is considerably lower than the mean age of 37 for video game
players overall~\cite{ESA:2011}. Of the 1191 respondents, an overwhelming
majority (94.9\%) reported their sex as male (compared with 80.8\% male in
MMO players~\cite{williams2008plays} and 58\% reported by the industry
overall~\cite{ESA:2011}). The reported age distribution of the respondents
is relatively smooth, with the exception of a large spike at 18 (see
Figure~\ref{fig:age_dist}), presumably caused by individuals younger than
18 misreporting their age so as to avoid the parental consent requirement.%
\footnote{The initial gamertag and age screening in the survey did not hint
  at the subsequent parental consent requirement for under 18s; thus,
  misreporting of age for this group may have been an automatic response to
  being queried for their age. However, participants could not be
  prohibited from refreshing the survey, reentering their gamertag and then
  misreporting their age.}

This contrasts with the reported age distribution of MMO games, in which
the plurality of players are in their 30s rather than their
20s~\cite{williams2008plays}.  However, the age distribution also exhibits
a long right tail, with 12.8\% being at least 30 years old. In agreement
with the age distribution, the most frequent response indicated some
college education.  Although 35 countries were represented in the
respondents, nearly three-quarters of respondents were from the United
States, with another 14\% from Canada and the United Kingdom.  An
overwhelming majority---94\% of our respondents---indicated English was
their first language.

Our respondents were very active players: they reported playing video
games for an average of 23.3 hours per week. Although this number may
appear high to non-gamers, it is slightly lower than the 25.9 hours per
week reported by MMO players and the 27.5 hours per week reported by the
industry in 2007 for video game players in
general~\cite{williams2008plays}.

\subsection{Psychometrics} \label{sec:psycho}

\begin{figure}[t!]
\begin{center}
\includegraphics[width=0.95\columnwidth]{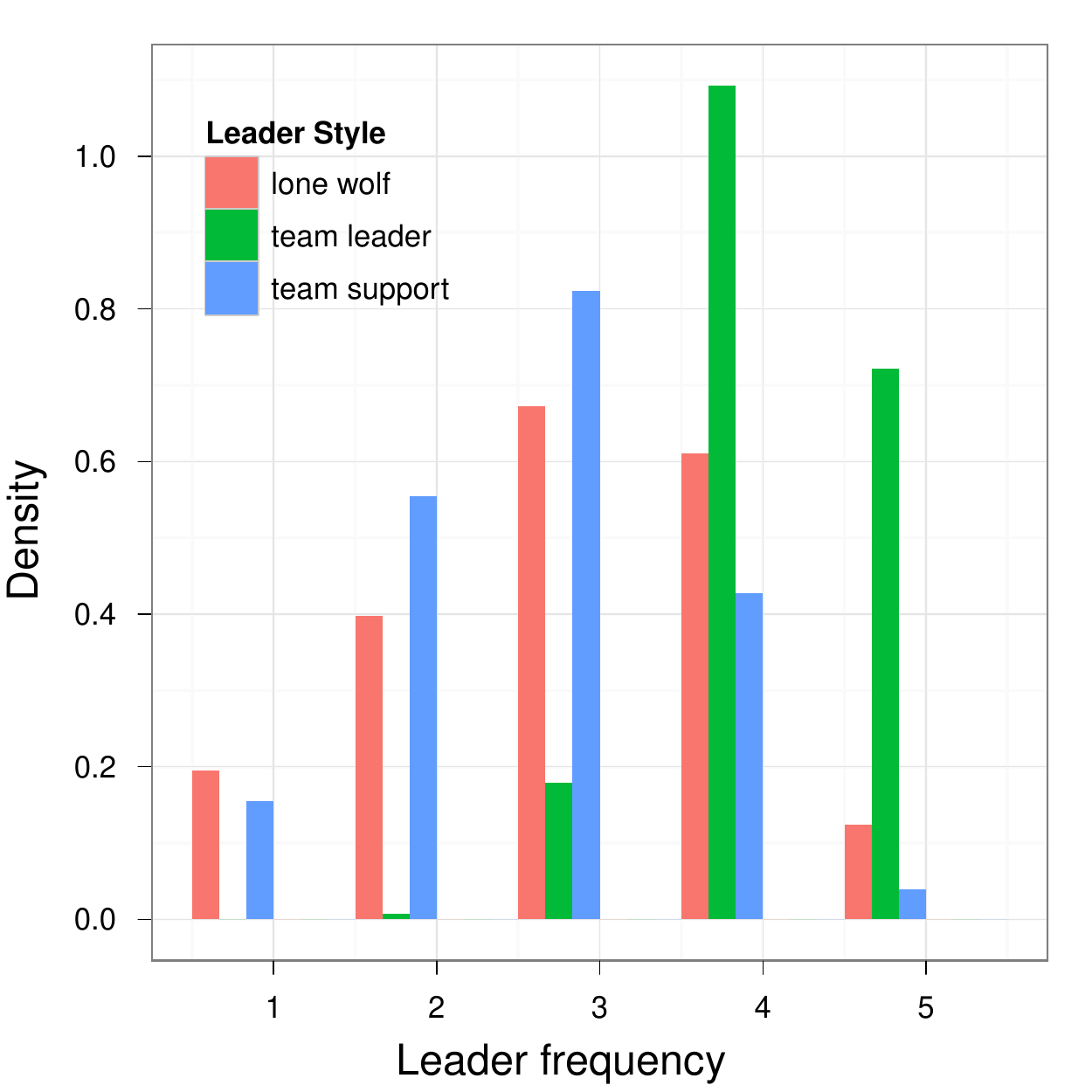}
\end{center}
\vspace*{-3mm}
\caption{Reported frequency of playing a leader role on a 1--5 scale where
  1 is never and 5 is always, by the preferred playing style.
\label{fig:ldrfreq}}
\vspace*{-3mm}
\end{figure}

We also asked participants to report their preferred playing style: team
leader or team support (both collaborative styles) or ``lone wolf'' (a style that does not coordinate
actions with the team).  The survey did not provide definitions of these
labels to respondents; however, feedback from our expert players indicated that their meanings should have been commonly understood within the \textit{Reach} community.
Although the popular stereotype of FPS games is non-collaborative, we find that 78.6\% of players prefer playing as a team, in either the leading or supporting roles.
Thus, {\em Reach} players are in fact strongly motivated by the
collaborative aspects of the game~\cite{xu:etal:2011,
  steinkuehler:williams:2006}.  When asked how often they played as a
leader, participants reported typically playing their preferred role,
although interestingly the lone wolves report playing a leader much more
often than we would naively expect, given the anti-social nature of the
lone wolf style (Fig.~\ref{fig:ldrfreq}).

We find some support for the popular perception that younger players prefer
the non-collaborative lone wolf role~\cite{wired}.  In our sample, people who preferred
the team support role were indeed older (23.5 years) than those who
preferred to play as lone wolves (21.6 years; $p < 0.001$). The modest size
of the difference, however, suggests that while the effect is real, it is
only a very weak tendency, with many younger players preferring collaborative
roles and many older players preferring non-collaborative ones. Interestingly,
people who preferred to play as leaders were also significantly younger (21
years) than those who played team support ($p \approx 0$).

Using our expertise as \textit{Reach} players, we modified three classic
psychometric surveys intended to assess group dynamics: entativity (how
much a group is like a single entity), cohesion (how tightly-knit a group
is), and conflict (how much internal conflict a group has). Past work on
collaboration and team performance suggests that low-levels of conflict
within a group of friends should correlate with better
performance~\cite{jehn:mannix:2001, shah:jehn:1993,jehn:shah:1997}.

The entativity portion of the survey contained 14 questions, which had
reasonable variability (SD=0.976 on a 5-point scale) and good internal
consistency (Cohen's $\alpha=0.82$). We therefore averaged the responses to
create a single entativity metric.  The average perceived entativity was
3.93 out of a 5-point scale, indicating respondents typically felt their
teams were good at acting as one. The portion of the survey on cohesiveness
contained 7 questions, which exhibited similar variability (SD=0.972) and
internal reliability ($\alpha=0.83$), so again we averaged the responses to
create a single metric.  In agreement with the perceived entativity,
players typically felt their teams were very cohesive (4.1 out of
5). Finally, the portion of the survey on group conflict consisted of 8
questions with slightly lower variability (SD=0.92) and less internal
reliability ($\alpha=0.76$), which appears to be related to the presence of
reverse-scored questions.  Nonetheless, we felt the consistency was
sufficient and averaged the scores.  Here, perceived conflict was low
(1.77), but somewhat higher than might be expected given the high
entativity and cohesion ratings.

\begin{figure}[t!]
\begin{center}
\includegraphics[width=0.95\columnwidth]{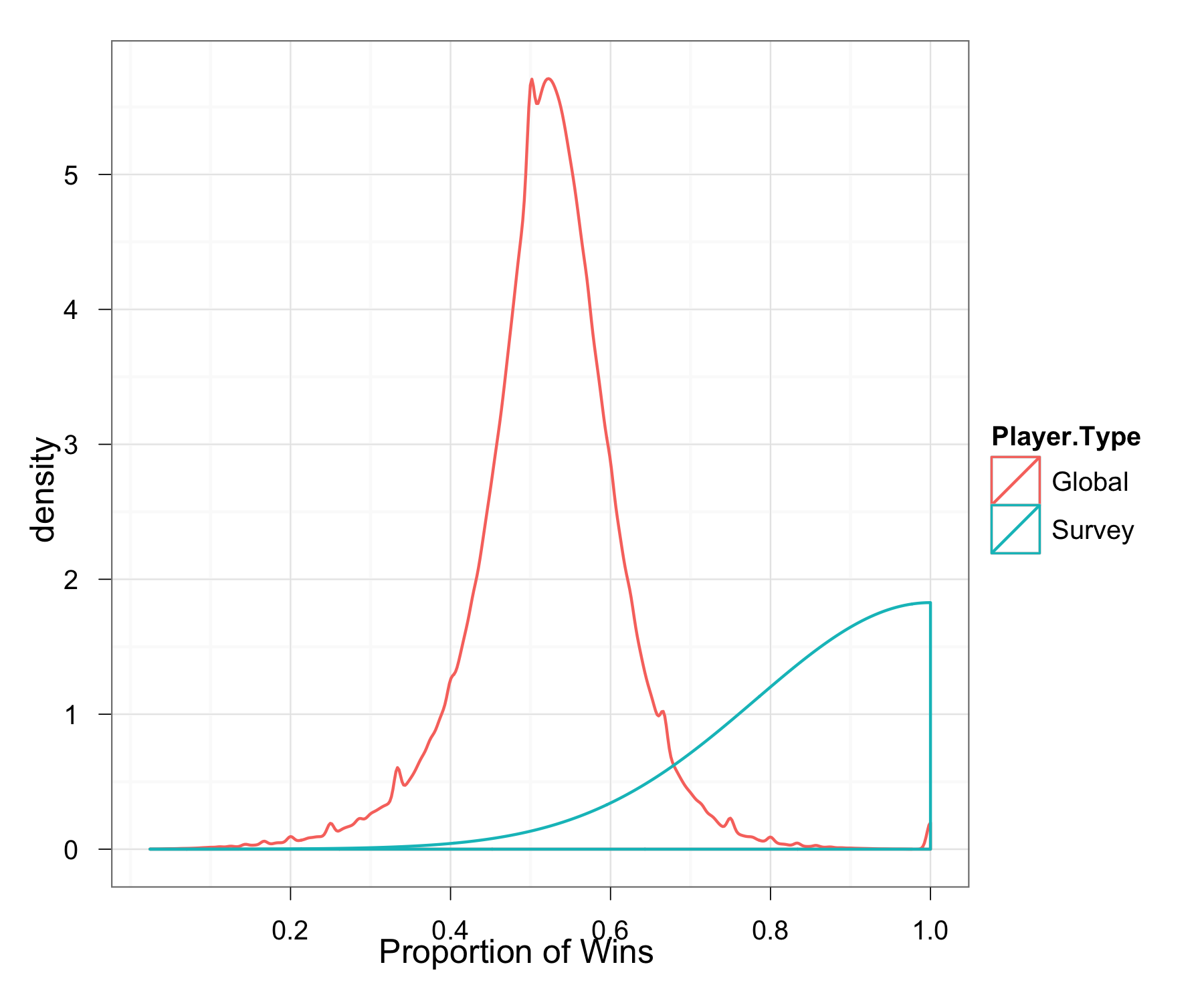}
\end{center}
\vspace*{-3mm}
\caption{Proportion of games won by our respondents compared to randomly selected players
\label{fig:wins_963}}
\vspace*{-3mm}
\end{figure}

\subsection{Comparison to Random Players}
\label{sec:963}

Advertising the anonymous survey on Halo-oriented community websites likely
produced a biased sample of respondents relative to the overall population
of {\em Reach} players. To characterize how different our respondent
population was from the overall player population, we downloaded a
uniformly random $1/1000$ sample of the first 400,000,000 {\em Reach} games
from which we extracted 963,000 unique gamertags.%
\footnote{The API does not provide a method by which to directly select a uniformly random player.}
We then downloaded each of their player detail files and estimated the
background distributions of various player summary statistics. Within this
population, the average proportion of wins exhibited by a player is almost
exactly 1/2, indicating a fairly unbiased sample of all players
(Fig.~\ref{fig:wins_963}).

\begin{figure}[t!]
\begin{center}
\includegraphics[width=0.95\columnwidth]{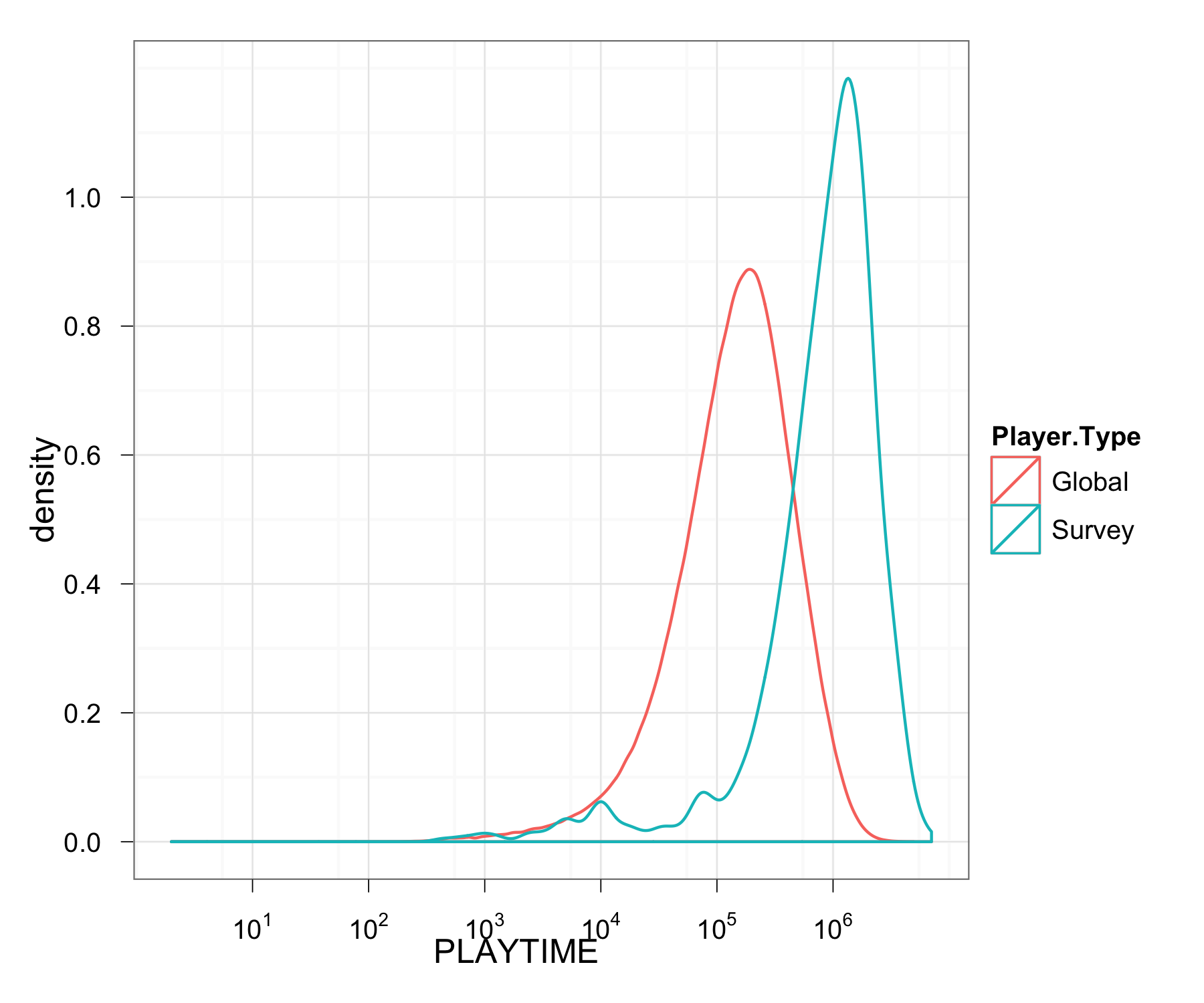}
\end{center}
\vspace*{-3mm}
\caption{Time spent playing {\em Halo:\!\! Reach} by survey participants versus randomly selected players
\label{fig:playtime_963}}
\vspace*{-3mm}
\end{figure}

Compared to the background distribution, our respondents were much better
players, winning many more of their games (Fig.~\ref{fig:wins_963}) than
the typical {\em Reach} player. In addition to being more successful
players, our respondents also typically invested almost 10 times more
actual time playing {\em Reach} than the typical player
(Fig.~\ref{fig:playtime_963}). That is, our respondents were serious Halo
players, likely devoting a large share of their overall time spent playing
video games to this particular game.

\subsection{Demographics and Game Play}
Although popular FPS stereotypes suggest that demographic variables like player
sex may correlate with game performance, we find that males and females
behave and perform similarly. For instance, males and females played
roughly the same number of games (Female $= 1825.2$, Male $= 2085.0$), and
the same amount of time per game (Female $= 592.3$, Male $= 620.69$; total
seconds played divided by total games played).  However, women died
slightly more often than men (Female $=10.69$ deaths/game, Male $=8.86$
deaths/game, $t(1189)=5.91$, $p<0.007$).%
\footnote{For all tests, the threshold for significance ($\alpha$) was
  corrected for the number of tests done.}
Of course, because of the relatively small number of female players (5.1\%
or 61 respondents), there may be other small but real effects that we are
unable to detect.

\begin{figure}[t!]
\begin{center}
\includegraphics[width=0.9\columnwidth]{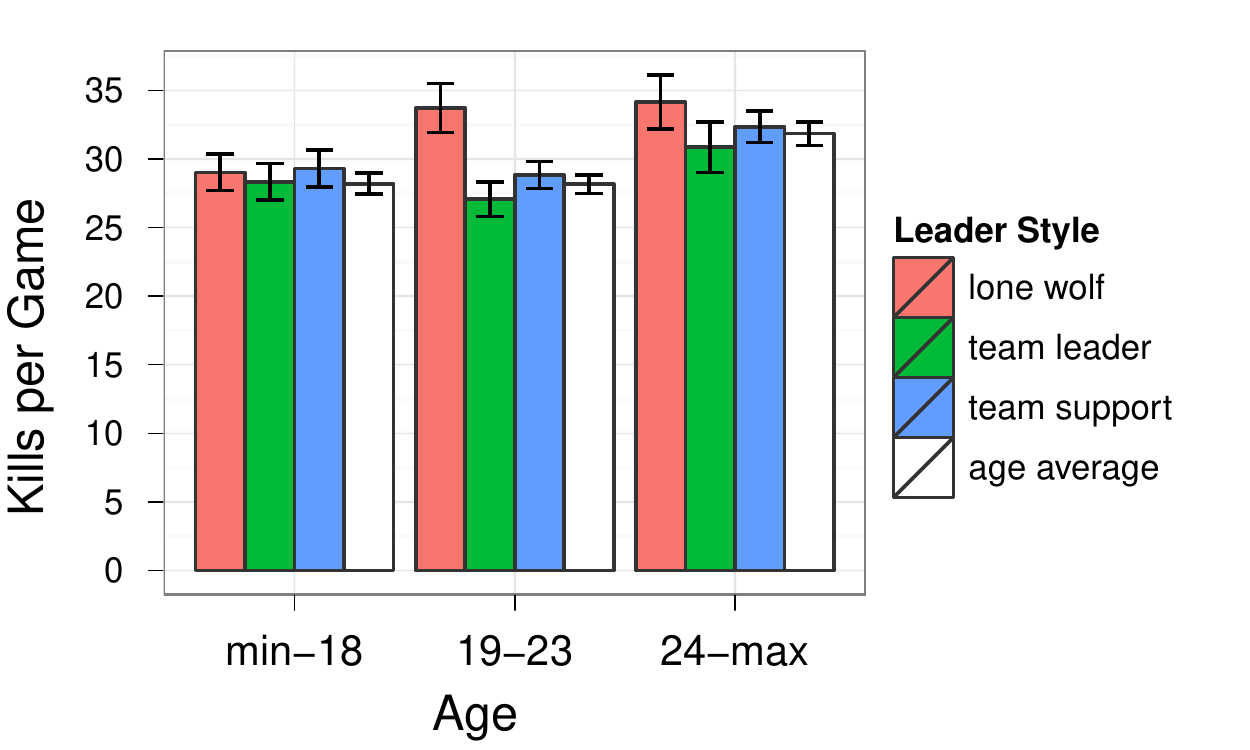} \\
\includegraphics[width=0.6\columnwidth]{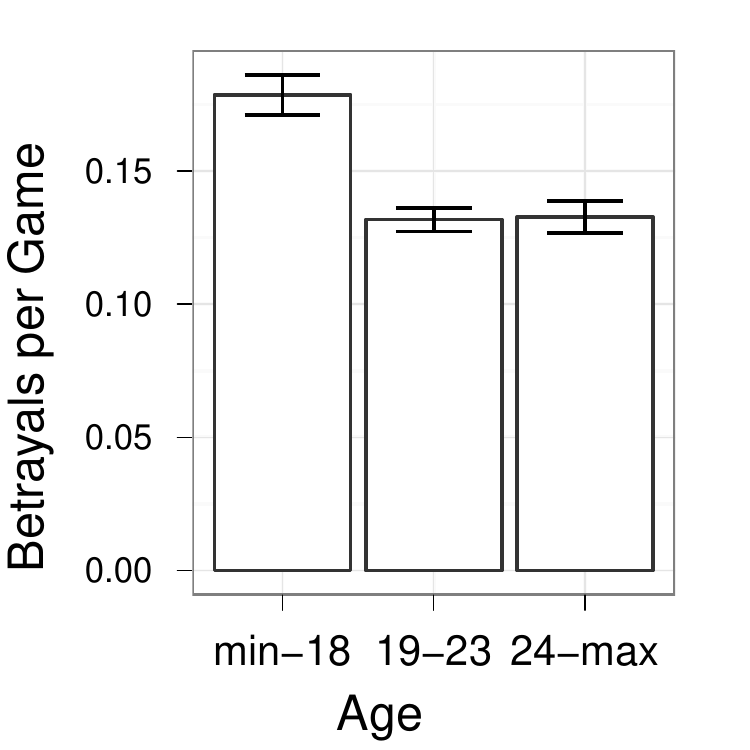}
\end{center}
\vspace*{-1mm}
\caption{Kills per game (top) and betrayals per game (bottom) by age group,
  showing that the oldest group has significantly more kills per game than
  the younger groups while the youngest has more betrayals per game than
  the older groups.
\label{fig:by_age}}
\vspace*{-3mm}
\end{figure}

We also observed interesting differences in game play by player age.
Congruent with the popular stereotypes of FPS gamers, the older a
respondent was the more kills he had per game ($\beta=0.263$,
$t(1189)=4.03$, $p\approx0$), as shown in Figure~\ref{fig:by_age}. We also
see older players were significantly more capable in the lone wolf role, a
result that runs contrary to the suggestion that young players prefer the
lone wolf role because they are more capable in the role---popularly
attributed to faster reflexes---while older players must coordinate in
teams in order to effectively compete~\cite{wired}.

One surprising age-correlated behavioral difference is in the number of
betrayals (killing one's own teammate, which results in a loss of a point
to the team and a longer respawn time for the offending player), as shown
in Figure~\ref{fig:by_age}.  Younger players (age $\leq$ 18) showed a
disproportionate amount of this type of team disloyalty relative to the
older players ($\beta=0.0024$, $t(1189)=-4.71$, $p\approx0$), with the
former group exhibiting a betrayal rate about 40\% higher than the latter
groups. Because betrayals result in a penalty against the team's overall
score, this suggests younger players exhibit significantly increased
anti-social behavior while older players generally exhibit a more
pro-social or cooperative orientation.

\subsection{Psychometrics and Game Play}
Unlike purely survey-based studies, the access granted through the API to
the detailed in-game data allows us to pair each survey response with rich
independent behavioral data on the same individuals. Through this pairing,
we may quantitatively verify reported variables on in-game behaviors. If
the survey responses are accurate, they should predict corresponding
patterns within the behavioral data, and allow us to subsequently predict
survey responses from behavioral data alone. For instance, participants who
report preferring the lone wolf role do indeed have significantly fewer
``assists'' to teammates (in which two or more team members collaborate to
score a point) than participants who prefer to be team leaders ($p \approx
0$) or team support ($p < 0.005$).

We also find meaningful differences in the reported group dynamic
psychometrics.  For instance, the more a player sees their group as an
entity, the more assists they have per game ($r=0.13$, $p \approx 0$).
Surprisingly, the opposite is true for number of kills: the more entitative
groups have significantly fewer kills per game ($r=-0.16$, $p \approx 0$),
perhaps because these groups travel as a pack and thus share their kills
among their number.  There are no significant differences in the number of
deaths per game, however, suggesting perhaps that such groups are no better
at warding off attacks than less entitative groups.

We observed a similar effect for perceptions of cohesiveness---more
cohesive groups had significantly fewer kills per game ($r=-0.1$,
$p<0.001$), and there were no differences in the number of deaths per game.
We did not, however, see the same relationship between cohesiveness and
assists that we did for entativity and assists, suggesting the measures may
be tapping different aspects of the players' subjective perceptions of
their groups.  A player going out of their way to assist a teammate, and
thereby potentially giving up opportunities to score their own points, may
require a greater sense of unity than mere closeness or camaraderie.

Studies of team dynamics in the management
literature~\cite{jehn:mannix:2001, shah:jehn:1993,jehn:shah:1997} suggest
that groups with greater levels of internal conflict should exhibit
different game play dynamics than groups with lower levels. Surprisingly,
however, we found no strong or notable relationships between responses
about group conflict and game play. There are several possible explanations
for this pattern, but we do not explore them here.

\section{Predicting Friendships} \label{sec:friends}

Before we consider the question of whether ground-truth friendship labels
can be predicted purely from in-game behavioral data, we briefly discuss
a few additional results from the survey data regarding friendships.

Of the 1191 respondents, 597 played at least one game with another survey respondent.%
\footnote{This overlap is likely attributable to the way we advertised the
  survey and is not representative of the background population.}
This overlap allows us to test whether respondents' perceptions of their
online and offline friendships are reciprocal~\cite{newman:2010}, i.e.,
when one player labels another as a friend, that friend also labels the
player as a friend. For online friendships, the reciprocity is 36.9\%; that
is, when one of two survey respondents indicated the other was an online
friend, about a third of the time did they agree they were friends. The
agreement on offline friendships was higher (60.9\%). While both of these
values may seem low, both are in fact much higher than the rates observed
in other online social networks, e.g., Twitter
(22\%)~\cite{kwak:etal:2010}.  For a social network where friendship ties
represent a significant mutual investment of time, it is not clear why we
do not see rates approaching 100\%, although there are a few possible
reasons.  For instance, we notice that there are a disproportionate number
of respondents who list exactly 29 online friends, and this could indicate
a number of respondents did not realize the list of teammates was
scrollable.  If this is the case, online and offline friends who fell
``under the fold'' may have been missed. Another explanation may be
variability in the interpretation of the term ``online friend'' or actual
differences in friendships of \emph{Reach} players.  This is a question for
future research.

\begin{figure}[t]
\begin{center}
\includegraphics[width=0.95\columnwidth]{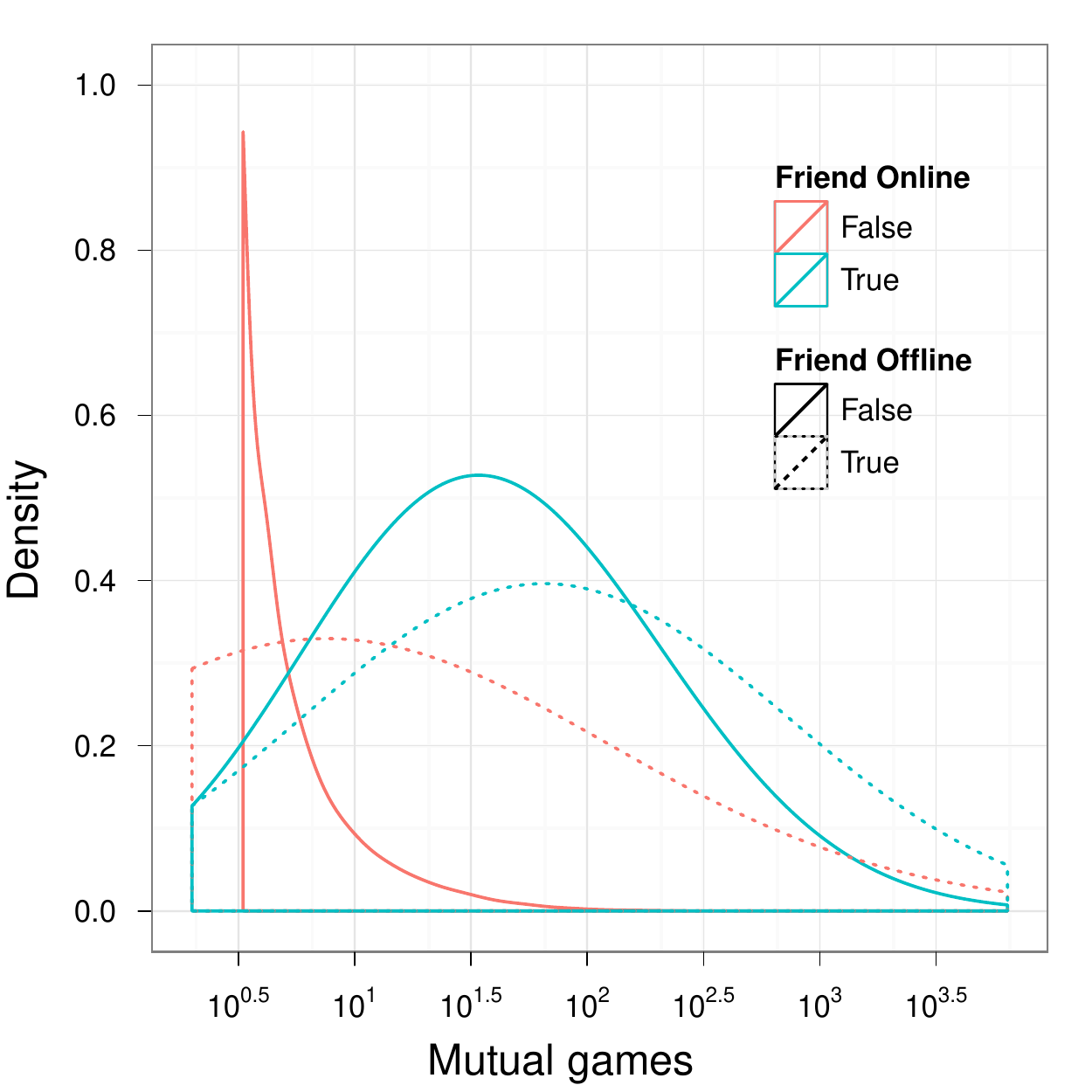}
\end{center}
\vspace*{-3mm}
\caption{Number of mutual games played with players declared as online and
  offline friends. Dashed lines represent offline friends, and blue lines
  represent online friends, so the solid blue line is an online friend who is
  not an offline friend.
\label{fig:friends_mutual}}
\end{figure}

\begin{figure*}[t]
\begin{center}
\includegraphics[width=0.75\textwidth]{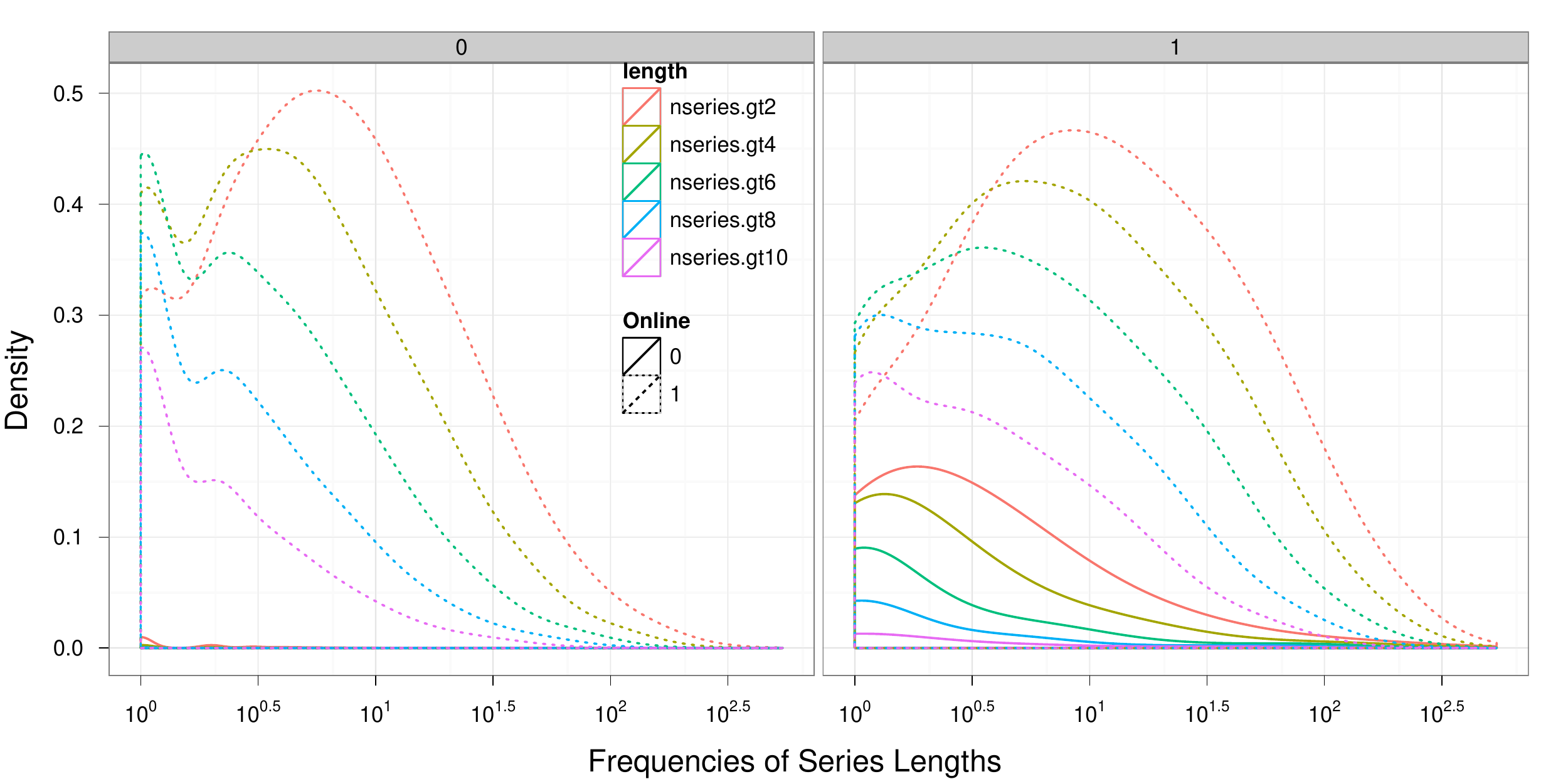}
\end{center}
\vspace*{-1mm}
\caption{Number of series played for online and offline friends.  Left
  panel shows series lengths for co-players who are not offline friends,
  right panel shows series lengths for offline friends.  Dashed lines are
  online friends, solid lines are co-players who are not online friends.
  Colored lines show distributions for different minimum number of games
  played in a series.
\label{fig:nseries}}
\end{figure*}

If players choose to structure their online activities in {\em Halo:\!\!
  Reach} around opportunities to play with friends, then the friendship
labels respondents provided to us should be predictable from in-game
behavioral data alone. Logistically, friends often synchronize their play
times, at the same time of day or day of the week; this leads each player's
gamertag to appear in the other player's game files in highly non-random
ways. Within \textit{Reach}, if friends are to play together, they must use
the in-game ``party'' mechanism, in which one player ``joins'' another
player to form a larger unit. Once they have formed this unit, the players
are automatically placed together by the matchmaking algorithm into the
same new competition and almost always on the same team (though splitting a
party can happen, which leads to interesting in-game behavior; see below.)
Thus, players who are friends will tend to appear together in a sequence of
games and, within each sequence, they will typically be on the same side of
a competition.%
\footnote{By default, {\em Reach} will party up all players within a
  particular game at its conclusion, after a 10 second delay. To prevent this, a 
  party or player must ``back out'' and restart the matchmaking
  algorithm. Most players choose to back out most of the
  time.}

We find significant differences in the frequency of playing together when a
participant labeled a co-player as an online friend, offline friend, both,
or neither. Pairs of players who are both online and offline friends
typically played the most together, followed by online but not offline
friends, then offline but not online friends and finally strangers
(Fig.~\ref{fig:friends_mutual}). That is, friends of both types play many
more games together than do strangers.

Additionally, the party mechanism implies that when
friends play together, their gamertags should appear together in a series
of games, and therefore these series should be a good predictor of
friendship.  We define a \emph{series} to be a sequence of consecutive
games played together with no more than a one-hour gap between
consecutive games. (Other gap sizes yield similar results to those reported
here, indicating the robustness of this feature.) For each pair of players,
we measured the number of series greater than length $n$ and the length of
the longest series.  Figure~\ref{fig:nseries} shows that online friends
were significantly more likely to have played series of games together, as
were offline friends. The greatest number of such series was played by
pairs who were both online and offline friends. Similarly, the longest
series for online and offline friend pairs is significantly longer than the
longest series for non-friend pairs.

To illustrate this point, consider the average longest series. For
strangers, the average was only 1.25 games in a row, indicating a strong
tendency for these groups to dissolve quickly. However, for pairs of
players in which either labeled the other as an online friend, the average
was 10.20 and for those labeled as both online and offline friends, it was
13.15 games (corresponding to roughly two hours of clock time). That is,
friends play many more games together in a sequence (on average, greater
than 8 times more) than do strangers, and the structure of the party
mechanism in {\em Reach} implies that this behavior must be actively
selected for by players.

These patterns suggest that certain purely behavioral features may be
highly accurate predictors of unknown friendship labels. Recovering this
kind of hidden information has been done with a range of other electronic
data on social interactions, including mobile phone call
records~\cite{eagle:clauset:quinn:2009,Eagle:2009tf}, geographic
coincidences~\cite{crandall:etal:2010} and email~\cite{Choudhury:2010vs},
but not previously interactions within online games.

\begin{figure}[t]
\begin{center}
\vspace*{-8mm}
\includegraphics[width=0.9\columnwidth]{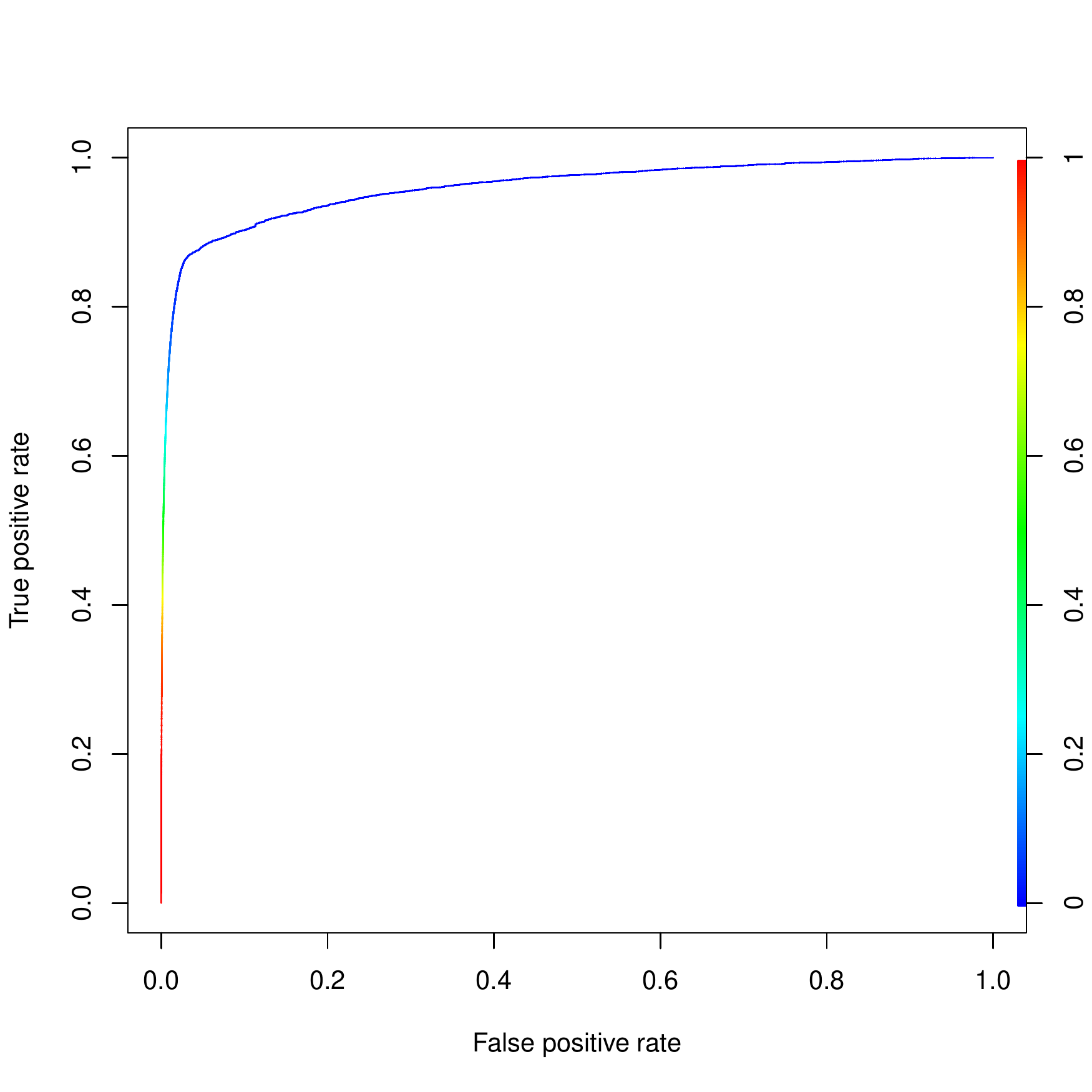}
\end{center}
\caption{ROC curve for predicting whether a player is considered an online
  friend, given the number of mutual games together, the longest series of
  games, and the number of series of different lengths.
\label{fig:pred_friends_on}}
\vspace*{-3mm}
\end{figure}

Given the number of competitions two players played and the number of those that
were played together, the length of the longest series, and how many series
of different length the pair played together (here: $\{2, 4, 6, 8, 10\}$,
although other values could also be used), we try to predict whether either
player indicated the other was an online friend (for those cases where
multiple teammates responded to the survey).  We fit the data to a logistic
regression model, and obtain an accuracy of 98.7\%, representing an AUC of
0.96. The corresponding ROC curve can be seen in
Figure~\ref{fig:pred_friends_on}, and the ranking of the features can be
seen in Table~\ref{tab:pred_friends_on}.

\begin{table}[b]
\begin{center}
\begin{tabular}{rrrrr}
  \hline
 & Estimate & Std. Error & z value & Pr($>$$|$z$|$) \\ 
  \hline
(Intercept) & -4.2485 & 0.0220 & -192.81 & 0.0000 \\ 
  longest.series & 0.3929 & 0.0030 & 131.80 & 0.0000 \\ 
  ngames.player & -0.0006 & 0.0000 & -73.96 & 0.0000 \\ 
  ngames.mutual & 0.0968 & 0.0027 & 35.97 & 0.0000 \\ 
  nseries.gt2 & -0.1686 & 0.0152 & -11.05 & 0.0000 \\ 
  nseries.gt4 & 0.0220 & 0.0197 & 1.11 & 0.2651 \\ 
  nseries.gt6 & -0.2075 & 0.0259 & -8.02 & 0.0000 \\ 
  nseries.gt8 & -0.3428 & 0.0331 & -10.36 & 0.0000 \\ 
  nseries.gt10 & -1.1609 & 0.0350 & -33.13 & 0.0000 \\ 
   \hline
\end{tabular}
\end{center}
\caption{Parameters in a logistic regression predicting whether a player is
  labeled an online friend}
\label{tab:pred_friends_on}
\end{table}

The dominant predictive feature in this model is clearly the length of the
longest series the two players played together.  The number of mutual games
is also a strong positive feature, though it's relevance is tempered by the
total number of games played.  The high accuracy of the model indicates
friendships in \emph{Reach} can be reliably inferred from online behavior
using simple heuristics. Given the fairly general nature of the predictive
features, we expect these results will generalize to other online
environments where players have the option of partying, or otherwise
preferentially playing with friends. As a caveat, our labeled friendships
are biased toward players with long game histories, and the most predictive
features leverage this deep behavioral data. To what degree friendships can
be inferred using only short behavioral histories is an important open
question.

\section{Friendship and Performance} \label{sec:winners}

Because we have all competition data for each player and the
participant-provided friendship labels of who they were playing with, we
can investigate whether a player performs quantitatively differently when
they play with or without friends. For the task of predicting winners based
on the number of friends on one's (or one's opponent's) team, we take two
simplifying steps. First, we focus on 4-on-4 games, which are the majority
of \textit{Reach} competitions; this controls for the number of players on
each team and eliminates the need to differentiate between number of
friends on a team or proportion of friends on a team.  Second, we aggregate
a player's online and offline friends; if a player labeled another as
either an online or an offline friend, we labeled that person as a
``friend''.

\begin{figure}[t]
\begin{center}
\includegraphics[width=0.8\columnwidth]{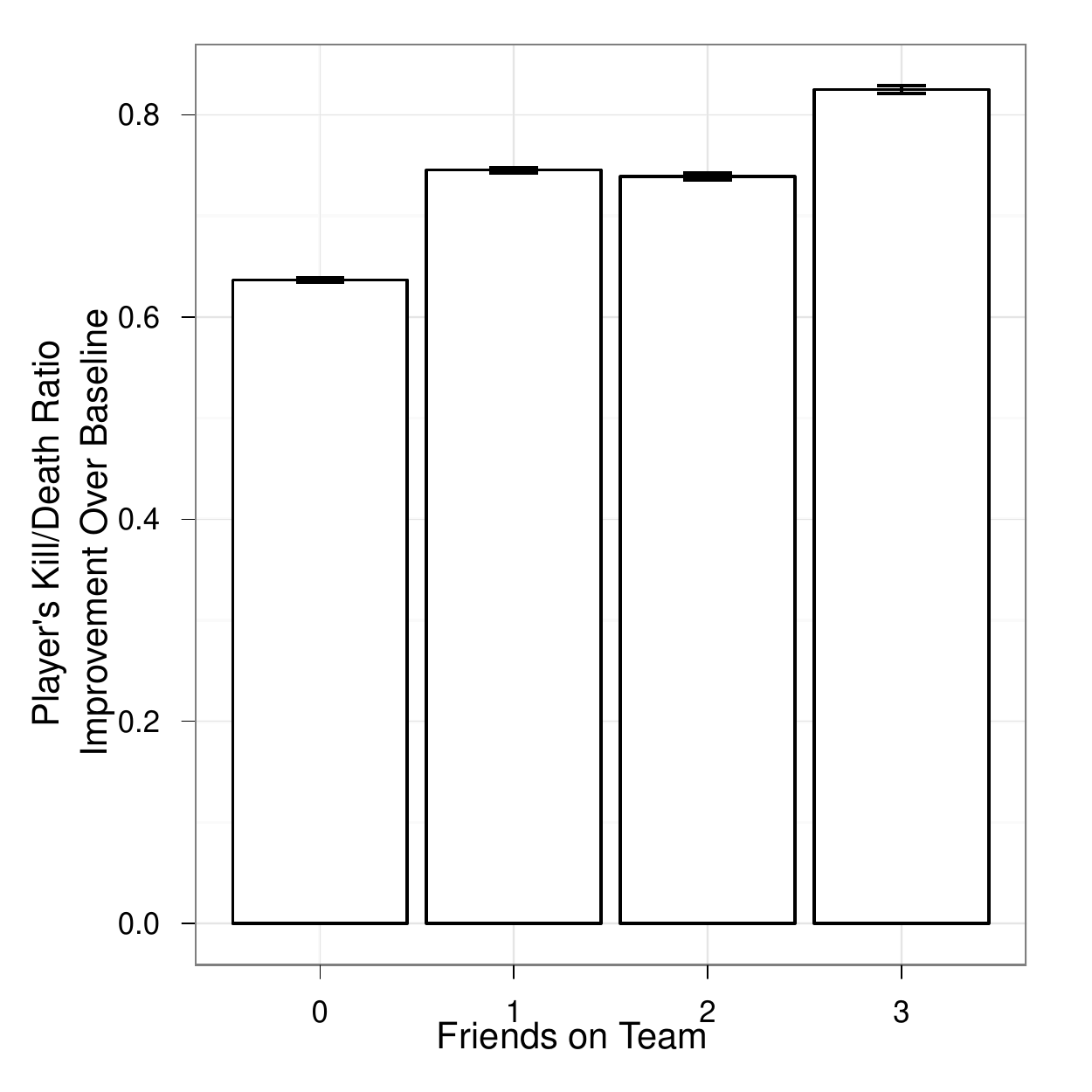}
\end{center}
\vspace*{-3mm}
\caption{Average kill-death ratio, for playing with 0-3 friends. Baseline
  is the population average kill-death ratio (1), so even with no friends
  on the team participants are 65\% better than
  average.} \label{fig:selfKD}
\vspace*{-3mm}
\end{figure}

\begin{table}[b]
\begin{center}
\begin{tabular}{rrrrr}
  \hline
 & Estimate & Std. Error & z value & Pr($>$$|$z$|$) \\ 
  \hline
(Intercept) & 0.2027 & 0.0031 & 64.47 & 0.0000 \\ 
  own team & 0.2558 & 0.0022 & 115.92 & 0.0000 \\ 
  opponent's & -0.1933 & 0.0069 & -28.16 & 0.0000 \\ 
   \hline
\end{tabular}
\end{center}
\caption{Best-fitting parameters for logistic regression model predicting
  whether a player won a game based only on the number of online or
  offline friends on their own team and on their opponent's team.}
\label{tab:wins_friends}
\end{table}

\subsection{Individual Performance}

Here we ask whether the presence of friends on one's own or on one's
opponent's team affects that player's performance.  In other words, do
people play or perform differently when they have friends on their team? If friendships impact the success of collaboration, then we expect to see performance improve as a player collaborates with more friends.

\begin{figure}[t]
\begin{center}
\subfigure{
\includegraphics[width=0.45\columnwidth]{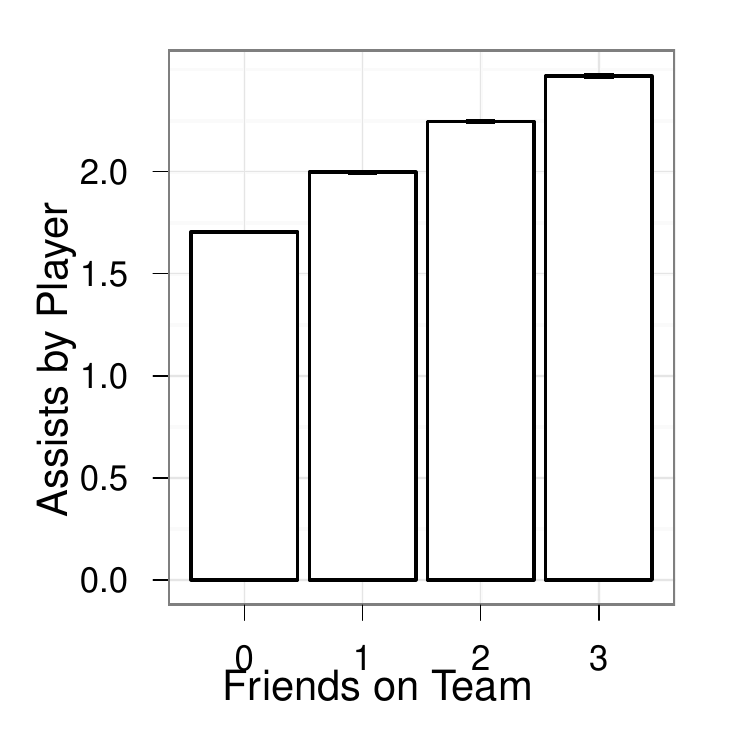}
}
\subfigure{
\includegraphics[width=0.45\columnwidth]{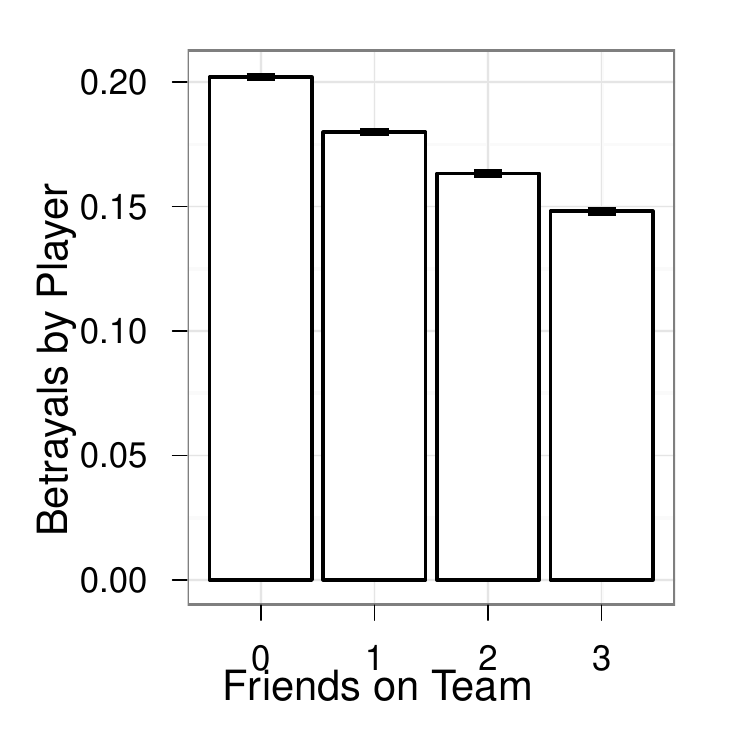}
}
\end{center}
\vspace*{-3mm}
\caption{Assist and betrayal frequency, for playing with 0-3 friends
\label{fig:assist_betrayal}}
\end{figure}

A strikingly clear result is that the amount of cooperation and defection
(measured as assists and betrayals) a player exhibits depends strongly on
the number of friends on their team.  Specifically, the more friends on
one's team, the more one assists (pro-social) and the less one betrays
(anti-social) one's teammates (Fig.~\ref{fig:assist_betrayal}). The
implication is that players actively adjust their play based on their
friendships---the motivation to maintain these relationships is greater
than the motivation to maintain harmony within the current team or to win
the current competition.

Playing with friends also impacts the player's success in the game, i.e.,
their performance.  Winning and losing a competition is a team outcome (see below), but a measure of individual performance is
the ratio of a given player's kill to deaths, i.e., the number of points
they personally scored versus the number of points they personally provided
to the opposing team. A higher individual value indicates better individual
performance; by definition, the population average kill-death ratio must be
1.0, as there is one death for every kill.  Figure~\ref{fig:selfKD} shows
that even when the survey respondents are playing with no friends on their
team, they do better than the average by 65\%, indicating that our survey
population is highly skilled.  However, this already strong performance further increases with
each additional friend on the team, up to an additional 20\% when the entire team is a group of friends.  Thus, at least
for our respondents, having friends on one's team has a real and direct
positive effect on one's own performance, independent of whether the team
goes on to win the competition.

\begin{figure}[t]
\begin{center}
\includegraphics[width=0.8\columnwidth]{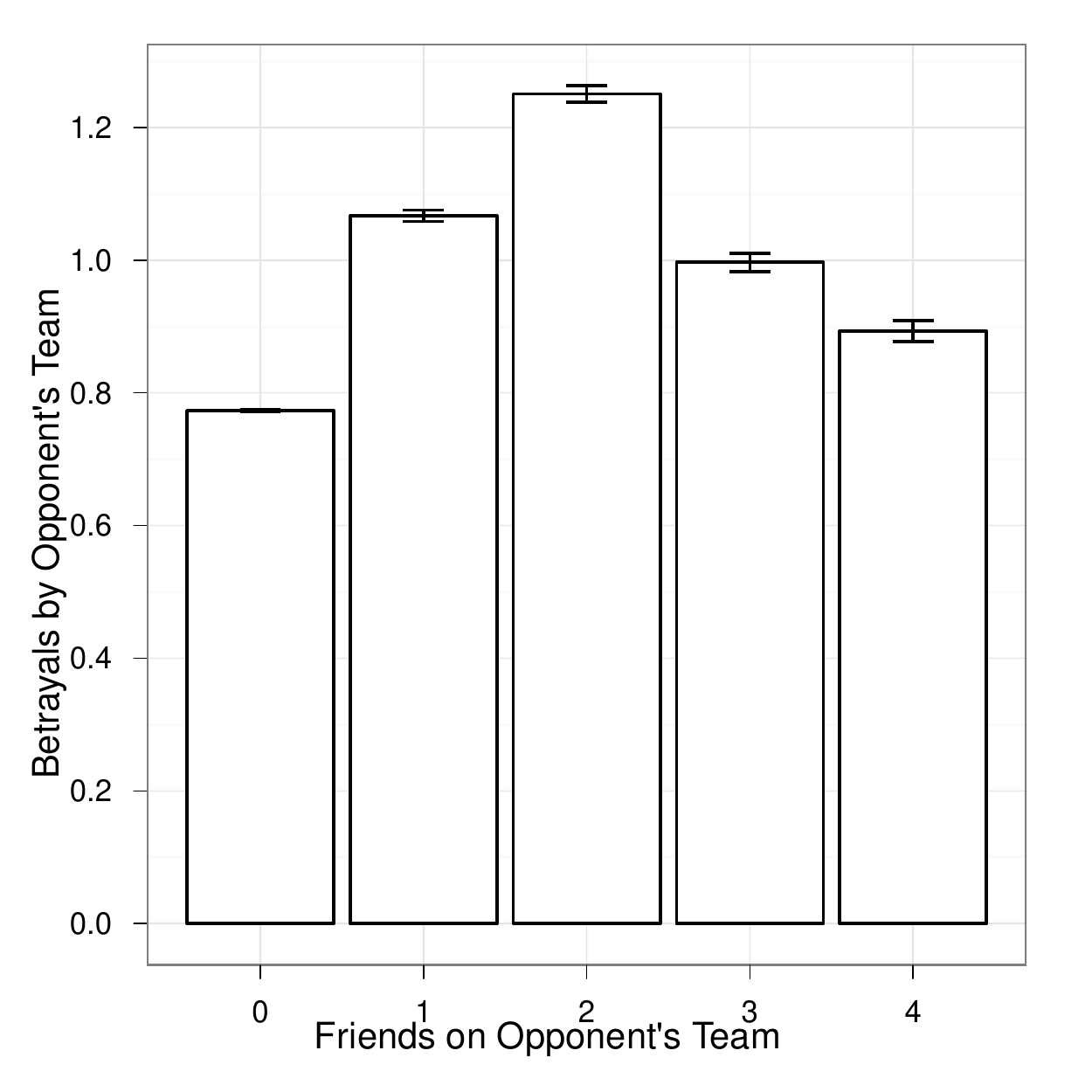}
\end{center}
\vspace*{-3mm}
\caption{The frequency of betrayals on opponent's team when the team
  consists 0-4 of the player's friends
\label{fig:betrayals_opp}}
\vspace*{-3mm}
\end{figure}

\subsection{Team Performance}

Given the effect of friends playing together on the individual's frequency
of assists and betrayals, it is not too surprising that we see a similar
pattern across the entire team.  That is, independent of the player's own
assists and betrayals, the team's assists increase, and its betrayals
decrease, with the number of friends playing together.

As mentioned above, the matchmaking algorithm sometimes places members of a party on the \textit{opposing} team, which allows us to investigate what happens when a player must choose between helping their teammates versus helping their friends (on the opposing team). In this case, we find an interesting pattern: as the number of friends on the opposing team
increases, so too does the number of betrayals on that team. This trend continues until
there is a majority of friends on the other team, at which point the number
of betrayals decreases again (Fig. ~\ref{fig:betrayals_opp}). In other
words, if one or two friends find themselves playing against their friends,
they are much more likely to kill their (non-friend) teammates than if they had no friends on the opposing team. The flip in the trend indicates that when a party's loyalties are divided across two teams, the smaller subset of friends is the one that defects against their teammates.

\begin{figure}[t]
\begin{center}
\includegraphics[width=0.8\columnwidth]{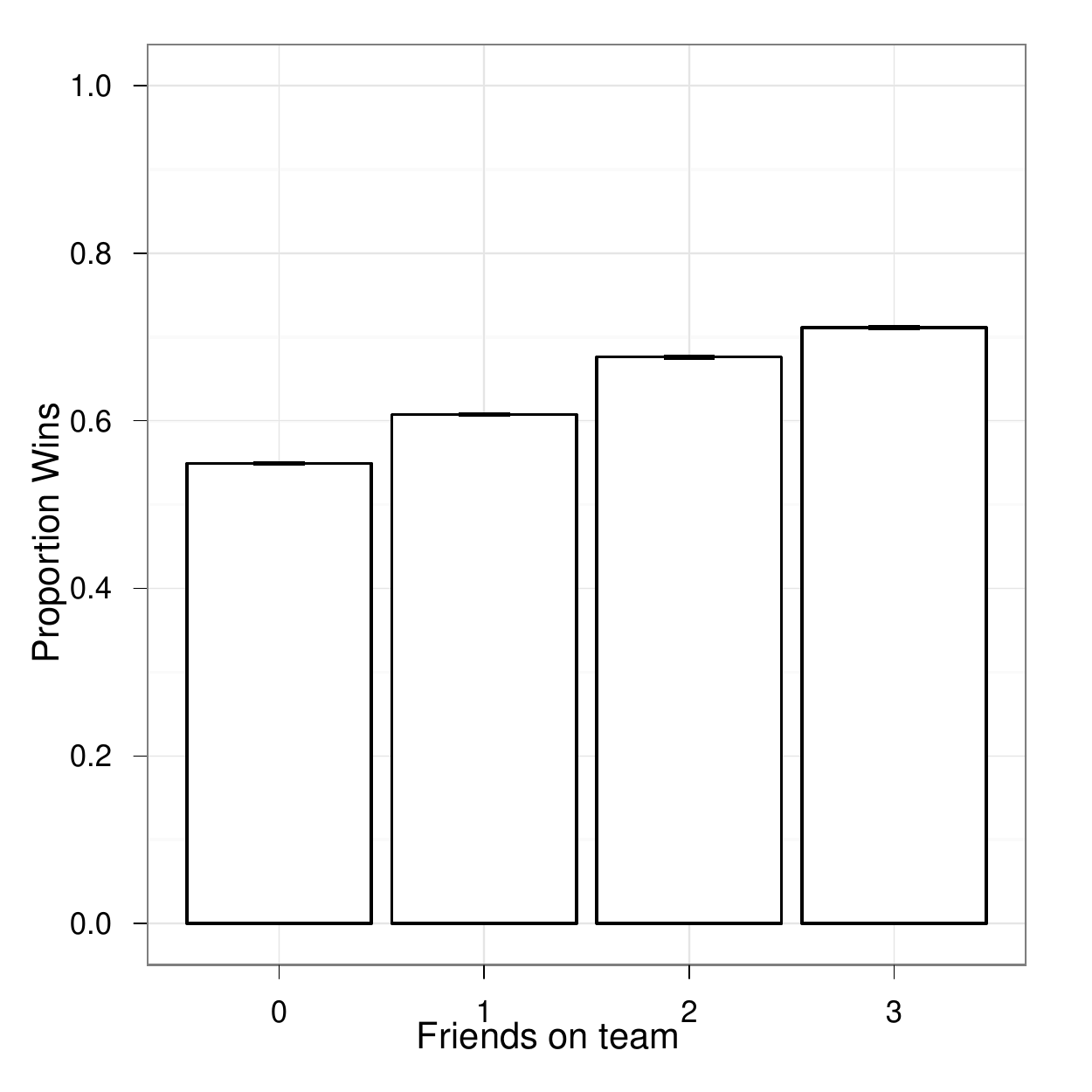}
\end{center}
\vspace*{-3mm}
\caption{Proportion of wins when playing with 0-3 friends
\label{fig:friends_wins}}
\vspace*{-3mm}
\end{figure}

Similar to prior experimental work~\cite{baldwin:etal:1997}, and in
agreement with our results on the kill-death ratio above, we find that
players win more often when playing with friends than with strangers, and
the more friends they play with the better (Fig.~\ref{fig:friends_wins}).
Notably, this effect appears despite the matchmaking algorithm's effort to
minimize differences in team skill~\cite{herbrich:minka:graepel:2007} and
thereby decrease the predictive power of non-game features like
friendship. That is, the TrueSkill matchmaking algorithm used by {\em
  Reach} attempts to control for the effect of skill in improving
performance, but it does not account for the performance benefits of from
friendship.

We fit a logisitic regression model using only information about how many
online and offline friends were on a player's own team and on the opposing
team.  The corresponding parameters are shown in
Table~\ref{tab:wins_friends}. The model achieves an accuracy of
61.1\%, with an AUC of 0.573.  We note that predicting the outcome of competitions using only the number of friends on a team is a very weak signal---being friends cannot make bad players good---so it is in fact highly meaningful that there is any improvement in accuracy over random chance (AUC of 0.5).

We also tried to predict a competition's winner based on a rough metric of
the players' expertise on each team. Because we do not have the entire game
history of \emph{all} players, we are unable to recreate the
TrueSkill~\cite{herbrich:minka:graepel:2007} estimates of the players'
skill level.  Moreover, despite the matchmaking algorithm's efforts to equalize skill
between teams, there is still considerable variability in skill levels
across competitions.  Instead we use reasonable proxies for a player's skill in
{\em Reach}, including the average number of games played by members of
each team, the average progress in the single-player campaign and
co-operative campaign, and the campaign difficulty level.  This model
performs slightly better than the friends-only model, improving the
accuracy to 62.2\% and the AUC to 0.614.  As can be seen in
Table~\ref{tab:wins_expertise}, more games played on one's own team is only
slightly more important than fewer games played by one's opponents.

Combining the expertise and friendship information provides a marginal
improvement over expertise alone, although both the level of expertise and
the number of friends on a team contribute to the model's prediction of the
outcome.  The three most heavily weighted parameters are the expertise on
one's own team, the opposing team's expertise, and the number of online
friends on one's own team (Table~\ref{tab:wins_all}).  The model has a
negligible improvement in accuracy, to 62.4\%, but the AUC improves
slightly to 0.625.

\section{Conclusion} \label{sec:conclusion}

In this paper we investigated the relationship of player demographic
variables, such as age, sex and group cohesion, with player behavior, and
the impact of friendships on individual and team performance in complex
competitive environments. Toward this end, we used the online game {\em
  Halo:\!\! Reach}, one of the most popular games of 2010~\cite{ESA:2011}
and one of the few in the FPS genre that has been studied in this way, as a
model system. Our results only scratch the surface of this large and
multi-faceted online system to inform our understanding of collaboration
and competition in virtual environments.

\begin{table}[t]
\begin{center}
\begin{tabular}{rrrrr}
  \hline
 & Estimate & Std. Error & z value & Pr($>$$|$z$|$) \\ 
  \hline
(Intercept) & 0.2578 & 0.0051 & 50.10 & 0.0000 \\ 
  own.ngames & 0.0001 & 0.0000 & 153.73 & 0.0000 \\ 
  oth.ngames & -0.0001 & 0.0000 & -136.39 & 0.0000 \\ 
   \hline
\end{tabular}
\end{center}
\caption{Parameters for the logisitic regression model predicting wins
  based on the average number of games played by one's own teammates and
  one's opponents.}
\label{tab:wins_expertise}
\end{table}

Our anonymous online survey produced a number of insights into the {\em
  Reach} community, many of which likely generalize to the FPS genre as a
whole. First, our respondents both invested more time in the game---but not
more time than a typical video game player spends on games of all
types---and were significantly more skilled than the typical {\em Reach}
player. Contradicting the anti-social stereotype of FPS players, our
respondents were strongly motivated by social factors. More than 99.5\% of
our respondents indicated that they played with at least one online friend,
and the typical number was 21 online friends (and 4 offline friends).%
\footnote{But, the degree distribution does not follow a power law~\cite{clauset:etal:2009}.}
Nearly 80\% preferred to play socially, on a team, rather than individually as a
``lone wolf,'' and most players felt a strong sense of group cohesiveness and
coordination. Thus, friendships among \textit{Reach} players play
an important role in choices of play style, game engagement and investment.

Although the typical age of our respondents was close to 20 years old, the
distribution shows a long tail, with a large number of players over 30. Age
correlated with several interesting aspects of game play: older players (24 or older;
30\%) tend to exhibit somewhat less within-group conflict, exhibit greater
pro-social tendencies (e.g., fewer betrayals), and are slightly more
skilled (more kills per game). This latter point is counterintuitive given
that younger players often have greater free time in which to invest in the
game. We found only small differences in play statistics between male and
female respondents and predictive models trained to predict sex from these
values scored no better than chance. Thus, male and female players seem
nearly indistinguishable from their in-game behavior.

Both team and individual performance in {\em Halo:\!\! Reach} are improved
by friendship variables and teams composed of friends win more games on
average than teams composed of strangers. However, if overall skill
correlates across friendship ties, then highly skilled groups of friends
could tend to win more often than groups of strangers because the skilled
friends are more skilled than an average stranger. That is, skill could be
homophilous.  However, the TrueSkill matchmaking algorithm, which actively
attempts to eliminate skill differences between teams, likely mitigates the
impact of this confounding factor.  TrueSkill does not control for the
benefits that comes from playing with friends---benefits we observe quite
clearly (Fig.~\ref{fig:friends_wins}). Admittedly, our study design does
not allow us to identify the social mechanism by which friendships improve
performance. A followup study that includes additional survey questions
designed to distinguish among candidate mechanisms, and perhaps in depth
interviews, may illuminate the answer.

\begin{table}[t!]
\begin{center}
\begin{tabular}{rrrrr}
  \hline
 & Estimate & Std. Error & z value & Pr($>$$|$z$|$) \\ 
  \hline
(Intercept) & 0.1740 & 0.0053 & 33.00 & 0.0000 \\ 
  own.friends & 0.1853 & 0.0023 & 78.99 & 0.0000 \\ 
  oth.friends & -0.0937 & 0.0071 & -13.26 & 0.0000 \\ 
  own.ngames & 0.0001 & 0.0000 & 125.61 & 0.0000 \\ 
  oth.ngames & -0.0001 & 0.0000 & -132.81 & 0.0000 \\ 
   \hline
\end{tabular}
\end{center}
\caption{Parameters for logistic regression predicting winners based on
  expertise and number of friends on one's own and one's opponent's team}
\label{tab:wins_all}
\end{table}

That being said, our results do suggest some interesting clues.
For instance, we observed more assists and fewer betrayals of teammates
when there were more friends on the team, in agreement with research
showing friendships decrease within-group conflict~\cite{jehn:mannix:2001}.
Alternatively, the large amount of time friends spend playing together may
improve the division of specialized roles and improve communication within
the team, leading to a better coordinated
effort~\cite{rittenbruch:mcewan:2009, jehn:shah:1997}.

Overall, friendships not only improve performance, but they also reshape the style of
play within competitions. That is, players compete differently when they
play with friends than when they play with strangers. For instance,
groups of friends who are split across teams have to choose whether to
compete against their friends (cooperate with their teammates) or defect
against the teammates (cooperate with their friends). When this happens,
friends tend to defect against their teammates, illustrated by a nearly
double betrayal rate when two friends are on the opposing team. That is,
friendship ties dominate teammate ties, and players sacrifice their own
competitive success to help their friends. Similarly, betrayal rates (both
by individuals and by teams) decrease as the number of friends on the team
increase, illustrating a significant pro-social effect as compared to teams
of strangers.

Although our logistic regression models found that friendship variables
make marginal improvements over skill variables in our ability to predict
success, this is understandable. In a context in which skill matters, a
team of unskilled friends is sure to do more poorly than a team of skilled
strangers. The important point is that even among skilled players,
friendship ties provide a lift in both individual and team performance.

To conclude, we found a number of interesting relationships between in-game
behavior within {\em Halo:\!\! Reach} and player demographic, group
covariates and friendship ties. Most generally, we found that friendships
strongly influence the social dynamics and collaboration within complex
virtual environments like the FPS game {\em Halo:\!\! Reach}. Friendships
lead to improved individual and team performance, increased pro-social
behavior, and likely increased long-term appeal of and engagement in the
game.

The impact of friendships is sufficiently strong, with players actively
structuring their in-game activities around opportunities to play with
friends, that hidden or unknown friendship ties can in fact be correctly
inferred directly from behavioral time series using common sense
heuristics. That is, it should be possible to automatically identify pairs
of individuals who would call each other friends, if asked, merely by
watching the way they interact with each other online. Collaborative
systems, including online FPS games like \textit{Reach} but also a wide
variety of other systems, that can infer and account for such friendship
ties could thus enable a wide range of better outcomes by providing fewer
mismatches between individuals or by suggesting individuals to interact
with in the future. Given the increasing ubiquity of social
gaming and {\em ad hoc} competition in general, the importance of such
algorithms is likely significant.

\section{Acknowlegements} 
We thank Chris Schenk for help implementing the web
survey, Nick Yee and Thore Graepel for helpful discussions, and the Halo
community for their support and participation. This project was funded in
part by a grant from the James S.\ McDonnell Foundation.

\end{document}